**Study of Counting Characteristics of Porous Radiation Detectors** *

M.P.Lorikyan

2 Brothers Alikhanian Str., Yerevan Physics Institute, 375036 Yerevan, ARMENIA

E-mail: lorikyan@moon.yerphi.am

lorikyan@star.yerphi.am

## Abstract

This paper presents the development of a new technology of registration of ionizing radiation and a new type of detectors – single-cathode multiwire porous detector with neither a gaseous nor semiconductor, but a porous dielectric substance, e.g., CsI, being used as working medium. It is shown that the performance of the multiwire porous detector is stable, ensuring highly efficient detection of both heavily ionizing particles and soft X-rays with a spatial resolution better than ±60μm. The continuous stable performance opens up new perspectives for using porous detectors in research as well as medicine. The obtained data are basic for the development of the theory of the phenomenon of electrons' drift and multiplication in porous dielectrics under the action of a strong external electric field.

* The work is supported by the International Science and Technology Center

The author is grateful to the founders of this Center for their huge support

## 1. Introduction

Porous detectors represent a new radiation detection technology and have a number of remarkable features. The use of porous dielectrics as working media in radiation detectors is a novalty in the radiation detection technique. Unique physical properties, such as stable performance for a long time, high spatial resolution (±60μm), good time resolution (≤60ps), low intrinsic noises, short pulses (≤1ps), high efficiency of registration of minimum and highly ionizing particles as well as X-rays and low cost allow them to be widely used in high-energy and nuclear physics experiments as well as in research, engineering and medicine. For example, low amount of material on the path of particles ($5\times10^{-4}$g/cm$^2$) and the possibility of operation in vacuum permit their use in time-of-flight devices and for beam monitoring in vacuum chambers of accelerators. Operation in vacuum allows them to be used also in space and other experiments. The technology of preparation of porous dielectric layers permits fabrication of detectors with an area of several hundreds of square centimeters. In case of using materials with high atomic number *Z*, one can ensure high efficiency of X-ray detection and usage of porous detectors as large area image detectors of X-rays. The detection efficiency of thermal neutrons in these detectors will be high if substances having a high neutron absorption cross-section, e.g., LiF, $GdCl_3$, are used as fillers. It is important that these detectors are comparatively low cost owing to the fact that the working materials to be used may be of not the highest purity (≤99.98%), and, unlike semiconductor detectors, their fabrication technology is simple.

Porous CsI, KCl, etc. are being used as working media in porous detectors. Operation of these detectors is based on the effect of collision ionization electrons' drift and multiplication in porous media under the action of an external electric field. The macroscopic mechanism of initiation of electrons' drift and multiplication (EDM) in porous dielectrics under the action of an external electric field and under the action of the field of charges having been accumulated on the surface of dielectrics (initiating an anomalous secondary electron emission (ASEE) (Molter Effect) [1-7]) is roughly the same – in both cases, there takes place an avalanche multiplication of electrons having been accelerated in the pores. In the second case, however, large space and surface charges in porous medium lead to substantial alterations in the physical processes and substantial divergences in the characteristics of these two phenomena.

The effect of ASEE within the range of KeV primary electrons has been experimentally studied quite well (see, e.g., ref. [5]). However, the large diversity of effects due to both unknown amount of impurities in the emitters being used and large space charge, make this problem a very complicated one, and, as far as the author is informed, there presently exists no comprehensive microscopic theory on this phenomenon. And what is more, such manner of initiation of an electric field makes ASEE inertial, non-controllable and unstable, and in case of minimum ionizing particles, the average secondary electron emission coefficient is low, $\mu$=1 [6,7]. That is why the experiments on using this phenomenon for radiation detection were a failure.

The effect of EDM in porous media under the action of an external electric field at bombardment by ionizing particles and soft X-rays was first observed and systematically studied by the group headed by the author in Yerevan Physics Institute in the early 1970ies [8-13]. To observe this effect, Lorikyan M.P, Trofimchuk N.N. et al. [8,9] prepared porous emitters of secondary electrons, which consisted of two fine-mesh grids with the gap between them filled with porous dielectric. A voltage was applied over the grids, and the emission characteristics were measured at the emitter's traversal by minimum ionizing particles. Fig.1 shows the number of secondary electrons emitted from porous layers of different thickness per primary electron as a function of the strength of electric field. It was shown that the process of electron emission was non-inertial, controllable and stable. The obtained results showed that the phenomenon could be used for particle detection.

Lorikyan M.P. and Trofimchuk N.N. created a porous detector with such an emitter [10]. That detector's efficiency for registration of $\alpha$-particles with ≈5MeV energy was 100%, and the average number of electrons emitted per one $\alpha$-particle was $n_e=270$, reaching $n_e=2\times10^3$ in ref. [14]. These results were confirmed in refs. [11-17].

On the basis of these investigations, the researchers of the same group created and studied multiwire porous detectors (MPD) of radiation [18-28]. It was shown, that the time resolution of these detectors was ≤60ps; the efficiency of registration of minimum ionizing particles and $\alpha$-particles with 5MeV energy, $\eta$=100%; and the coordinate resolution was better than 250μm. The main disadvantage of these detectors was that their performance was stable at a pulsed mode of operation, making them inefficient.

In the present work, porous multiwire detectors filled with CsI of 99.98 purity were developed and assembled at maximum possible cleanliness of all detector-manufacturing procedures. It was shown that such detectors performed stable at a constant voltage feed, i.e. at a continuous mode of operation, which means that the polarization effects manifested themselves weakly. Continuous-operation porous detectors strongly enhance the possibility of using porous detectors, and are a stimulus to the development of a quantitative theory of electrons' drift and multiplication in porous dielectrics under the action of electric field and ionizing radiation.

Drift and multiplication of electrons in dielectrics under the action of an external electric field has not yet been studied well even experimentally. Porous detectors of ionizing radiation are still under development.

For details in this field refer to ref. [26].

## 2. On the Principle of Operation of Porous Detectors

Note that the absence of a microscopic theory on effective drift and multiplication of collision ionization electrons in porous dielectrics under the action of an external electric field is seriously limiting the possibility for a quantitative interpretation of the obtained and prediction of the expected results, also is a serious hindrance to optimal conduction of experiments.

It is obvious, that efficient performance of porous detectors depends entirely on the number of primary-particle-induced δ-electrons and on the efficiency of δ-electron-induced secondary electron emission from pore walls.

Qualitatively the microscopic mechanism of EDM in porous dielectrics under the action of an external electric field can be represented as follows. In pore walls, primary particles knock out electrons from different energy bands of the dielectric, except for the conduction band, as it is vacant in dielectrics at the beginning. Part of these electrons in the same pore wall and part of electrons having penetrated into other pores and getting accelerated there produce electrons in other pore walls and transfer them into the conduction band. Electrons thus occupy the conduction band. In the conduction band, the electrons deposit energy mainly due to their interaction with phonons. Eventually, part of electrons occupy levels in the conduction band's bottom and higher, and part of electrons having energies higher than the electron affinity $\chi$ of the surface of the material escape from the pore walls into vacuum (into pores). The electrons are accelerated in the pores under the applied electric field and the same processes occur on the walls of other pores. These processes occur with all generations of secondary electrons, and if $\chi$ is small, the average coefficient of secondary electron emission in each act of electron collision with pore walls $\sigma>1$, then an avalanche multiplication of electrons takes place in the porous medium.

Under regular conditions $\chi$ is high for dielectrics, thermal electrons do not take part in the process of emission, and initiation of cascade processes of electron multiplication is practically impossible. But $\chi$ may significantly drop in materials with P-type conduction under the action of an electric field, because the electric field pushes the acceptor-induced holes from the near-surface regions of the walls of pores situated on the positive electrode side deeper into the bulk, forming a depletion zone in the close vicinity of that surface with an excess of negative charges (acceptors) in it. The field induced by these charges pushes the conduction band electrons towards the surface of pore partitions and facilitates their escape, i.e. $\chi$ drops. Decrease of $\chi$ results in the fact, that thermal electrons also contribute to the emission process, significantly outnumbering the hot electrons. Decrease of $\chi$ increases also the thickness of the layer (escape depth) from which these electrons can reach the surface and escape into vacuum, as thermal electrons are depositing little energy [28]. All these effects secure a high coefficient of secondary electron emission from every pore wall and efficient drift and multiplication of electrons in porous medium.

In case of anomalous secondary electron emission (ASEE), the processes described above take place in the presence of large surface and space charges (as ASEE arises under the influence of these charges) which keep continuously changing. So, drift and multiplication of electrons takes place in the presence of a large number of capturing centers, and the electrons' drift and multiplication becomes inertial and non-stable owing to the fact, that the number of these charges is in a direct dependence on the factor of multiplication of electrons.

Dielectrics have a high specific conductance. This is good from the viewpoint of intrinsic noises of porous detectors, but the usage of dielectrics as working media faces a problem connected with the defects in the crystalline structure, which are charge-carrier capturing centers and are responsible for the formation of a large space charge. The direction of the space charge's electric field is opposite to that of the external field, that is why the efficiency of electrons' drift and multiplication decreases, and in the course of time the drift of charge carriers stops, i.e. the polarization effects are undoubtedly affecting the counting response of MPD. Space charge itself is also an additional trap for charge carriers. Loss of charge carriers reduces the amplitude and increases the fluctuations of detector pulses. Slowing-down of the charge carriers' drift to the electrodes introduces additional fluctuations in the carrier accumulation time, i.e. deteriorates the detector's time resolution.

Besides, under the action of the summary field of external electrodes and space charges, the electrons get a transverse momentum, deteriorating the detector's spatial resolution. And what is more, the space charge grows in the course of operation of detector, changing the electric field in the working medium and deteriorating the conditions of electrons' drift and multiplication. So, detectors with dielectric working media having a notable density of carrier-capturing centers will have a poor spatial, energy and time resolution and unstable performance. These effects have been already observed in earlier refs. [18-27]. Beside this kind of traps in materials, there exist also impurities/traps, that is why the working medium's purity is so crucial for normal operation of porous detectors [29].

In ionization radiation detectors with porous working media made of dielectrics, the loss of charge carriers is compensated by multiplication of electrons on pore partitions. However, one should not expect from porous dielectrics an energy resolution comparable with the energy resolution of semiconductor detectors, as capturing of charge-carriers and vigorous electron multiplication introduce fluctuations in the height of the detector pulse.

## 3. The Experimental Method

The MPD being tested had porous CsI of 99.98–99.99 purity. The porous CsI layer was prepared by means of thermal deposition of CsI in an Ar atmosphere [40]. Porous CsI was deposited on the cathode. The thickness and density of layer were a function of deposition time, pressure of Ar, distance from the cathode to the vessel *l* from which CsI was evaporated, and temperature of the vessel. In our case, CsI was melted in a tantalum vessel. A porous CsI layer with a density of about 0.4% of single crystal density was deposited at an Ar pressure of $p$=3Torr, $l$=6.5cm, and a deposition rate of 43μm/min. A density of 0.75% was obtained at $p$=3Torr, $l$=5cm. The detector was assembled in the atmosphere of Ar immediately after CsI was deposited. After assembly, the detector was placed in a pressurized container filled with Ar and moved into a vacuum chamber. The measurements were taken under (7–9)×$10^{-3}$Torr vacuum. All detector manufacturing procedures were conducted under maximum possible technological clinliness.

To estimate the coordinate resolution, the anode wires of MPD were divided into two groups. The first group included the even numbered anodes and the second group - the odd numbered ones. The anodes in each group were connected to the inputs of fast current amplifiers having a rise time of 1.5ns and a conversion ratio of 30mV/μA. After amplification, the signals from the first and second groups were shaped and counted. The shaped pulses from each group were simultaneously fed to a coincidence circuit, by means of which the number of events $N_c$, when the same particle was detected by both groups, i.e., by two neighboring anodes (because the number of random coincidences was negligibly small), was also readout. In case of $N_c \neq 0$, it is obvious, that there is a strip in between two neighboring anodes, where the particles incident on it are registered by both anodes. The width of this strip, $\Delta x$, determines the detector's coordinate resolution. Denoting the spacing between the anodes by $b$, and the total number of particles registered by both anode wire groups by $N_0$, we have:

$$\Delta x / b = N_c / N_0,. \qquad (1)$$

and the number of particles registered by only the first or second group of anodes will be equal to $N(I)=n(I)-N_c/2$ and $N(II)=n(II)-N_c/2$ respectively, where $n(I)$ and $n(II)$ denote the number of pulses at the output of odd- and even-numbered anode wire groups, respectively, i.e. the number of particles the detection of which induced shaped pulses at the output of the first and second anode wire groups. The total number of particles registered by the detector will be *N(I)+N(II).*

The time of the measurement of each point was one minute. In the pauses between the measurements the detectors were exposed to detected particles and a voltage. When detectors were turned off, air evacuation from the vacuum chamber stopped, leaving the detectors in residual vacuum.

The dependence of *n(I), n(II),* $N_c$, *N(I)* and *N(II)* on *U,* and the time-stability function of MPD were investigated.

The errors in all cases are taken to be statistical and are not indicated. Intrinsic noises were ≈0.1-0.2$s^{-1}$.

## 4. Description of Detector

The schematic view of MPD is shown in Fig.2.

Anode wires *1* are stretched across the surface of fiberglass laminate frame *2*. Porous CsI layer *6* is first deposited on cathode *4* made of 30-100μm-thick Al foil. Frame *2* is fastened to the cathode with porous CsI deposited on it. The Al foil is stretched across frame *5* made of fiberglass. Dielectric frame *3* serves for setting of the desired gap between the cathode and anode. The thickness of deposited CsI layer, H, is larger than the gap between the cathode and anode wires, so that the anode wires are buried in the porous layer.

In the MPD tested, the anode wires are made of gilded tungsten wire having a diameter of 25μm, the gap being 0.5mm. X-rays were incident onto the porous CsI through a thin Al cathode, while α-particles got in the porous medium from the side of anode wires.

## 5. Results of Measurements

It was observed, that the thickness of porous CsI layer decreased after its thermal deposition on the support plate, i.e. there took place compaction of the porous layer. Fig.3 shows how the thickness of porous CsI decreases in the course of time after its deposition on the support plate. The observed compaction of CsI layer is accompanied by crystallization of the deposited material and formation of a polycrystalline structure, which leads to a decreasing number and density of bulk and surface defects that play a major role in the process of electrons' drift and multiplication and drift of holes under the action of electric field.

Figs. 4, 5 and 6 present the results of measurements taken by MPD-1 with a porous CsI layer having a density of $\rho\approx0.3\%$ and a thickness of H=0.8±0.05mm after its deposition, respectively. Anode wires were spaced at $b$=250μm. The sensitive area was $22\times22cm^2$. The detection threshold was $V_{thr}$=35mV. The total number of α-particles incident on the whole detector was $N_{0\alpha}=2562min^{-1}$.

Fig.4 shows the dependence of the number of α-particles detected by the first group $n_\alpha(I)$ (crosses) and second group $n_\alpha(II)$ (points) of anode wires on voltage $U$ in an hour after putting MPD-1 together. Triangles show the number of coincidence pulses $N_c$ from the first and second groups of anode wires, i.e. the number of cases when the same α-particle is detected by both groups of anode wires.

One can see that all the dependences have a plateau, and within the errors each group of wires is counting all α-particles, and the number of coincidence pulses $N_c$ is equal to the number of α-particles registered by each group of anode wires $n_\alpha(I)$ and $n_\alpha(II)$. This result indicates that all α-particles are in parallel registered by both neighboring wires, i.e. the spatial resolution of MPD is worse than the distance between the anode wires. So, the efficiency of particle detection by each half of the detector is expressed as $\eta=n_\alpha(I)/N_0(I)=n_\alpha(II)/N_0(II)=N_c/N_0$, where $N_0(I)$ and $N_0(II)$ are the number of α-particles incident on each half of the detector. On the plateaus $\eta$=99%.

The time-stability function of MPD-1 measured immediately after these measurements (Fig.4), is shown in Fig.5. It is seen that the detector's performance is stable for some time, but then $N_\alpha$, i.e. the detection efficiency, rapidly drops to zero and remains unchanged, and in doing so MPD stays spatially insensitive all the time ($N_c=n_\alpha(I)=n_\alpha(II)$).

After finishing these measurements (which lasted about 7 hours), we measured the dependence of $n_\alpha(I)$ and $n_\alpha(II)$ on $U$. The results are presented in Fig.6. The comparison of Fig.4 with Fig.6 shows that in Fig.6, first, all curves shifted to the right by ≈125V, second, $n_\alpha(I)$ and $n_\alpha(II)$ are twice as smaller and, third, $N_c$ has drastically decreased, i.e. after working about 7 hours, each group of anode wires had been detecting particles separately, which means that the detector became spatially sensitive. Fig.6 also shows that the MPD's spatial resolution decreased with increasing $U$. The real number of particles detected in this case by, e.g., only one group of anode wires is $N_\alpha(I)=n_\alpha(I)-N_c/2$, labeled by squares in the same Figure. The time-stability measurements showed that the MPD's performance was stable with a 100% efficiency of registration by each group. After this, the MPD's counting characteristics remained unchanged.

Fig.7 shows the results of measurements of $U$-dependence of $n_\alpha(I)$ (crosses), $n_\alpha(II)$ (points) and $N_c$ (triangles) by analogous MPD-2 with $\rho$=0.4% and H=1.3μm. The measurements were conducted at $N_{0\alpha}$=2800min$^{-1}$ and $V_{thr}$=26mV, in one hour after deposition of porous CsI. Again, as in the previous case, $n_\alpha(I)=n_\alpha(II)=N_c$, i.e. the detector was spatially insensitive. The measurements show that the time-stability function is the same as in the previous case (Fig.5).

Fig.8 shows the results of the measurement of the dependence of $n_\alpha(I)$, $n_\alpha(II)$, $N_\alpha(I)=n(I)-N_c/2$ and $N_c$ on $U$, taken after 14 hours' long rest of MPD-2 in vacuum, after obtaining the results presented in Fig.7. In this case also $n_\alpha(I)=n_\alpha(II)$, but both, $n_\alpha(I)$ and $n_\alpha(II)$ are twice as smaller, and $N_c$ has dropped dramatically, i.e. each group of anode wires, after putting the detector at rest, has acquired spatial sensitivity. Fig.9 shows the time-stability function of, e.g., $n_\alpha(I)$ (crosses), $N_\alpha(I)$ (squares) and $N_c$ (triangles) during 5 hours of continuous operation, measured after obtaining the results presented in Fig.8.

So, the investigations show, that there is indeed observed a significant change in the porous layer's characteristics on passage of a certain time after deposition of porous CsI layer, irrespective whether MPD is operated or turned off.

These results show, that the detection efficiency of MPD-2 drops (Fig.5) due to increasing of the working voltage in the course of time, as a result of which the plateau shifts to the right with a low $\eta$ at the previous values of $U$. Increasing of $U$ is conditioned by the formation of microcrystals and a polycrystalline structure in the course of time (Fig.3), resulting in compaction of the porous layer, shortening of the free path of electrons, and strengthening of the electric field necessary for the acceleration of electrons to energies high enough for initiation of avalanches. The transverse size of these avalanches also decreases, improving the detector’s spatial resolution.

Note, that in the first hours of operation, when MPD-2 was spatially insensitive, the curves of $n_\alpha(I)=f(U)$ and $n_\alpha(II)=f(U)$ are showing a well-expressed plateau (Fig.4). And when MPD-2 resumed spatial sensitivity, the plateaus on these curves become narrower (crosses and points) (Fig.6). This is because in the first case each anode wire is registering α-particles having hit mainly its sensitive area and some number of α-particles having hit the sensitive area of the neighboring wire ($N_c \neq 0$). The probability of registering particles having hit the neighboring anode wire increases with $U$ (Fig.6, triangles), that is why none of the curves can have a plateau. But the analogous dependences of the number of α-particles having hit the sensitive area of only one group of anode wires, $N_\alpha(I)$ and $N_\alpha(II)$ reaches a plateau. This is well seen in Fig.6, where the dependence of $N_\alpha(I)$ on $U$ is labeled by squares. This is understandable, because now α-particles having hit the sensitive area of the neighboring wire are discarded.

MPD-2 was operated 10 hours a day followed by a 14 hours' rest during 16 days, after which it was disassembled. Every day, the obtained data were practically the same as the results obtained on the second day (Fig.8), i.e. the detector's properties did not change after its getting stabilized.

The time-stability function of this detector having operated at $U$=1350V for 16 days is shown in Fig.10. Plotted along the abscissa are the days and along the ordinate – the average number of $N_\alpha(I)$ (squares) and $N_c$ (triangles) per diem. During this series of measurements MPD-2 had been put at rest on the 4th day for 38 hours, and on the 11th day for 60 hours. It also follows from these results, that the substantially long absence of exposure to radiation as well as the supply voltage do not affect the stability of MPD operation.

Similar investigations have been carried out also with MPDs with anode wires spaced at $b$=0.125mm and $b$=0.75mm. In case of $b$=0.125mm, the results were analogous to those mentioned above, and $N_c$ on the next day was equal to ≈6% of the total number of detected α-particles at a detection efficiency of 0.9, i.e. the MPD's coordinate resolution was much better than ±60μm. In case of $b$=750μm, $N_c$ was approximately 65% from the very beginning, while on the second day it was ≈0.2%.

There was also studied MPD-3 with an area of 5×5cm$^2$, the detector's gap and the spacing between the anode wires being 0.5mm, and a CsI layer with H=0.7mm and $\rho$=0.4%. The measurements of the thickness of CsI layer over the whole area showed that it is uniform within ±25μm. The number of incident α-particles was $N_{0\alpha}$=4320min$^{-1}$. This detector performed exactly like the previous ones. Fig.11 shows the results of the time-stability measurements of the detection efficiency conducted for 48 hours of continuous operation of MPD-3, beginning on the 3rd day after deposition of porous CsI layer. Fig.12 shows the results of the time-stability measurements conducted for 42 days of continuous operation of the same detector. In the latter case, the detector operated 8 hours a day, being switched off for the rest of the day. The fact that the detection efficiency reached 95% indicates that it is within an accuracy of 5% the same over the whole area of the detector.

Analogous measurements have been conducted on MPD-4 designed for X-ray detection with a porous CsI layer with H=1.3±0.05mm, $b$=0.5mm and $\rho\approx0.4\%$ after deposition. The intensity of X-rays was $74s^{-1}$, and the registration threshold $V_{thr}$=35 mV. The results of measurements show, that, again, as in the former cases, in the beginning MPD-4 is spatially insensitive, and the detection efficiency sharply decreases in the course of time. These results are shown in Fig.13 and Fig.14.

In Fig.13 are presented the dependence of $n_x(I)$ (crosses), $n_x(II)$ (points) and $N_c$ (triangles) on $U$. Fig.14 shows the time-stability function of MPD-4 obtained in one hour after the measurements shown in Fig.13.

The dependence of $n_x(I)$ (crosses), $n_x(II)$ (points), $N_c$ (triangles), and $N_x(I)=n_x(I)-N_c/2$ (squares) on $U$ measured after 16 hours' rest of MPD-4 and the time-stability data are presented in Fig.15 and Fig.16, respectively. One can see, that after taking rest, as in case of α-particles, MPD-4 acquires spatial sensitivity, time stability, and the working voltage increases. But, as distinct from the case with α-particles, the plateau on the curve of $N_x(II)=f(U)$ lost former clearness, and $\eta$ dropped a little faster (30% in 4 hours, Fig.16) than in case of α-particles. The investigations showed, that decrease of $\eta$ slowed down day by day, and on the 11th day it was 10% for 8 hours, $\eta$ having been dropping mainly during the first 2 hours, as was the case with α-particles.

However, as distinct from the case with α-particles, the X-ray detection efficiency, $\eta_x$, measured in one hour after deposition of CsI, was substantially lower than $\eta_x$ measured after the rest period, as in the latter case the number of X-quanta detected in each group of anode wires did not decrease as it did in case of α-particles, but increased. Apparently this is due to the fact, that the particle detection efficiency is a function of the ionizing power of particles. As far as in the beginning of operation the particle detection efficiency drops in the course of time (Fig.5 and Fig.14), it could be supposed that it keeps dropping also when measuring $n_\alpha(I)$ and $n_\alpha(II)$ as a function of $U$, forming a 'false' plateau. The ionizing power of α-particles is so high that electron loss due to decreasing efficiency of electrons' drift and multiplication does not substantially affect $\eta_\alpha$. As for X-rays, their ionizing power is ~500 times lower than that of α-particles, so that the decrease of the number of drifting electrons notably decreases $\eta_x$. After rest, the performance of MPD-4 is stable, that is why $n_x(I)$ and $n_x(II)$ do not decrease as a function of $U$, and there is no 'false' plateau.

The measurements by MPD-4 lasted 29 days. Sixteen days the detector was registering X-rays, 15 days of which the detector was turned off after having operated for 8-9 hours every day. On the 16$^{th}$ and 17$^{th}$ days the measurements were conducted round the clock, i.e. continuously, then the measurements were conducted in the previous mode for the rest of the time.

Fig.17 shows the time-stability function of MPD-4 measured at $U$=1550V on the 16$^{th}$ and 17$^{th}$ days for one of the anode wire groups, e.g., second group, $N_x(II)=n_x(II)–N_c/2$. The last four values of $N_x(II)$ were obtained at higher $U$ ($U$=1580V). Although the ionizing power of X-rays is ~500 times higher than that of α-particles, however, as in case of α-particles, the detector shows good time stability, and the slight decrease of $\eta$ is controllable.

To study the performance stability of MPD-4, on the 18$^{th}$ day after finishing the time-stability measurements, we once again measured the dependence of $N_x(II)$ and $N_c$ on $U$ (Fig.18). It is seen in Fig.18, that the working characteristics of MPD have not practically changed as compared with the second day (Fig.15), the maximum efficiency of X-ray detection being 65%. The X-ray measurements continued for another day with $\eta_x$ remaining the same. Then the measurements continued for 11 days, detecting α-particles. The detector performed stable with $\eta_\alpha$=100%.

It should be noted, that every day, in the beginning of the measurements, the number of coincidence pulses from two groups of anode wires was higher than at the end of the measurements, which means, that the MPDs' coordinate resolution was improving with the operation time.

So, MPDs perform stable at registering α-particles as well as X-rays with ≈6KeV energy and have a high spatial resolution.

In all the cases, in between the measurements the detectors were exposed to radiation and a voltage was applied over them.

On completion of the series of these measurements, all MPDs were disassembled, revealing no changes in the porous CsI layer, i.e. the detectors could have been still operable.

## 6. Discussion

Unfortunately, the experimental investigations in this case are going ahead of the theory and there is no possibility to conduct experiments based on theory predictions as was the case with, e.g., X-ray transition radiation detectors [30,31]∗. However, even qualitative representations show, that the reason for the absence of spatial sensitivity and stability of MPD performance at the beginning is the high density of bulk and surface defects in the porous medium. The increase of porous layer's density in the course of time is obviously accompanied by a decrease in the number of these defects, so that the probability of capturing charge-carriers decreases, and, correspondingly, the space and surface charges also decrease. So, MPDs get stabilized and acquire higher spatial sensitivity. The latter is affected also by the shortening of the transverse size of electron avalanche.

It should be noted that it is possible to obtain better results by using CsI of higher purity, ensuring cleanliness of the technological procedures of fabrication of the whole of the detectors, and taking of certain measures, as, e.g., increasing the number of acceptors and coating the pore walls by Cs and oxygen.

## 7. Conclusions

The results of the investigations testify to the fact that the MPD characteristics strongly change on passage of some time after deposition of porous CsI layer.  We presume that in freshly prepared CsI layers deposited by thermal evaporation, in the course of time there takes place spontaneous formation of microcrystals and a polycrystalline structure. This results in decreasing of the density of charge-carrier-capturing centers which slows down the

* As was shown in [32], X-ray transition radiation was not detected in earlier works [33,34].

formation of space and surface charges, and the MPDs get stabilized and acquire spatial sensitivity.

So, multiwire porous detectors, conserving their spatial sensitivity, are maintaining a long-term stability of performance, and the insignificant drop of the particle detection efficiency can be corrected by adjusting the supply voltage.

It is also obvious, that the experimental data are insufficient for a final quantitative interpretation of the discovered effects.

The obtained results are basic, and besides their practical significance, they will stimulate the development of the theory of electrons' drift and multiplication in porous dielectrics under the action of a strong electric field and ionizing radiation, will broaden and deepen the knowledge on solids, and open new perspectives for using dielectrics in such fields of engineering that are non-traditional for using dielectrics.

**Acknowledgements**

The author is thankful to Ajvazyan G., Ajvazyan R., Vardanyan H., Asryan G., Papyan G. and Aghababyan G. for assistance in preparation of this work.

## Figure captions

**Fig.1** The average number of secondary electrons emitted from porous KCl layer versus the strength of electric field at different thicknesses of porous layer.

**Fig.2** The schematic view of MPD: 1 – anode wires, 2 – fiber glass laminate frame, 3 – dielectric frame, 4 – cathode, 5 – supporting frame, 6 – porous CsI layer.

**Fig.3** Change of the thickness of porous CsI in the course of time after its deposition on the support plate.

**Fig.4** The dependence of the number of α-particles detected by the first group $n_{\alpha}(I)$ (crosses) and second group $n_{\alpha}(II)$ (points) of anode wires and the number of coincidence pulses from the first and second groups of anode wires $N_c$ (triangles) on voltage, in one hour after putting the MPD-1 together.

**Fig.5** The time-stability function of MPD-1 obtained immediately after the measurements shown in Fig.4. Crosses – $n_{\alpha}(I)$, points – $n_{\alpha}(II)$, and triangles – $N_c$.

**Fig.6** The dependence of $n_{\alpha}(I)$ (crosses), $n_{\alpha}(II)$ (points), $N_c$ (triangles) and $N_{\alpha}(II)$ (squares) on $U$ for MPD-1 measured after obtaining the data shown in Fig.5.

**Fig.7** The dependence of $n_{\alpha}(I)$ (crosses), $n_{\alpha}(II)$ (points) and $N_c$ (triangles) on voltage, in one hour after putting MPD-2 together.

**Fig.8** The dependence of $n_{\alpha}(I)$ (crosses), $n_{\alpha}(II)$ (points), $N_c$ (triangles) and $N_{\alpha}(II)$ (squares) on $U$ measured after 14 hours' long rest of MPD-2, after obtaining the data shown in Fig.7.

**Fig.9** The time-stability function of $n_{\alpha}(I)$ (crosses), $N_c$ (triangles) and $N_{\alpha}(II)$ (squares) measured after finishing the measurements of Fig.8.

**Fig.10** The time-stability of MPD-2 at $U$=1350V during 16 days. Squares – $N_\alpha(II)$ and triangles – $N_c$.

**Fig.11** The time stability data acquired during two days of operation of MPD-3 measured in 3 days after CsI deposition, $N_x(I)$ – squares.

**Fig.12** The time stability data acquired during 42 days of operation of MPD-3, squares – $N_\alpha(I)$.

**Fig.13** The dependence of $n_x(I)$ (crosses), $n_x(II)$ (points) and $N_c$ (triangles) on voltage, in one hour after putting MPD-4 together.

**Fig.14** The time-stability function of MPD-4 obtained immediately after the measurements shown in Fig.13. Crosses - $n_x(I)$, points – $n_x(II)$ and triangles –$N_c$.

**Fig.15** The dependence of $n_x(I)$ (crosses), $n_x(II)$ (points), $N_c$ (triangles) and $N_x(I)$ (squares) on $U$ measured after 16 hours' rest of MPD-4, after obtaining the results presented in Fig.14.

**Fig.16** The time-stability function of $n_x(I)$ (crosses), $n_x(II)$ (points), $N_c$ (triangles) and $N_x(I)$ (squares) obtained after finishing the measurements of Fig.15.

**Fig.17** The time stability of MPD-4 during 52 hours of continuous operation at $U$=1550V. In the last four hours of measurements $U$=1585V. Squares – $N_x(II)$ and triangles – $N_c$.

**Fig.18** Dependence of $N_x(II)$ (squares) and $N_c$ (triangles) on $U$, measured on the 18$^{th}$ day of operation.

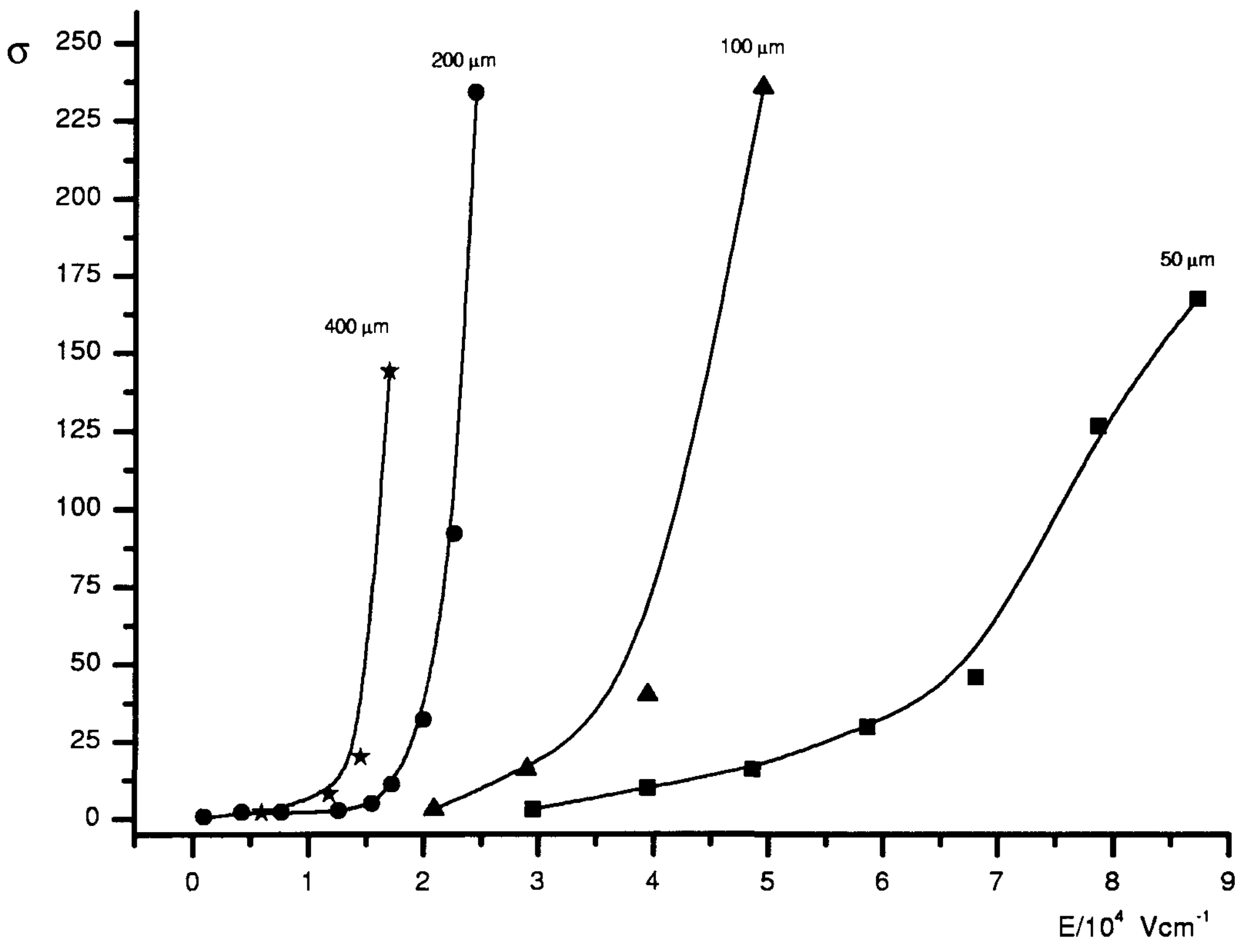

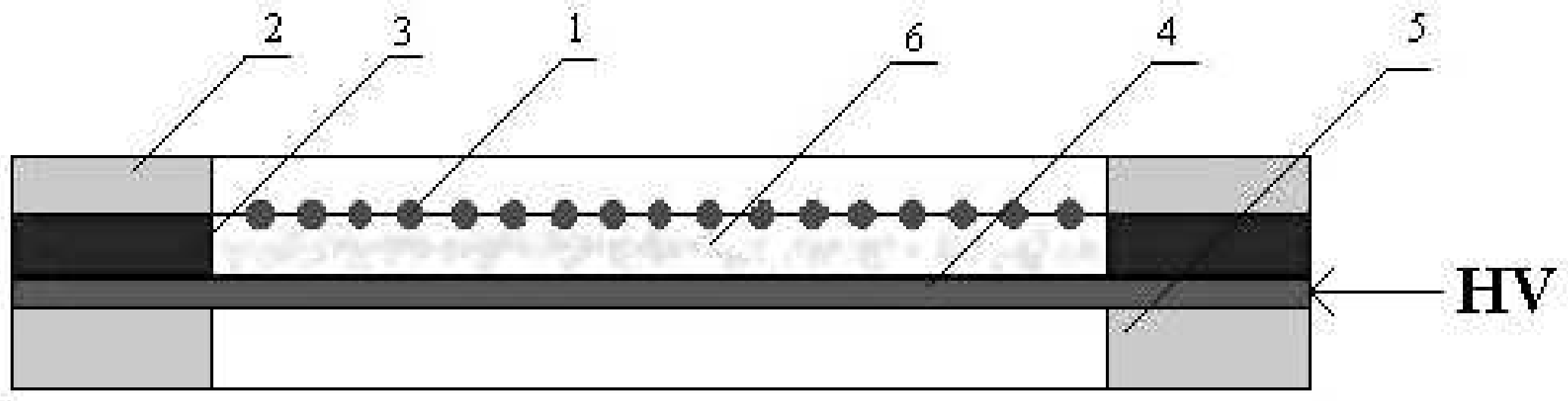

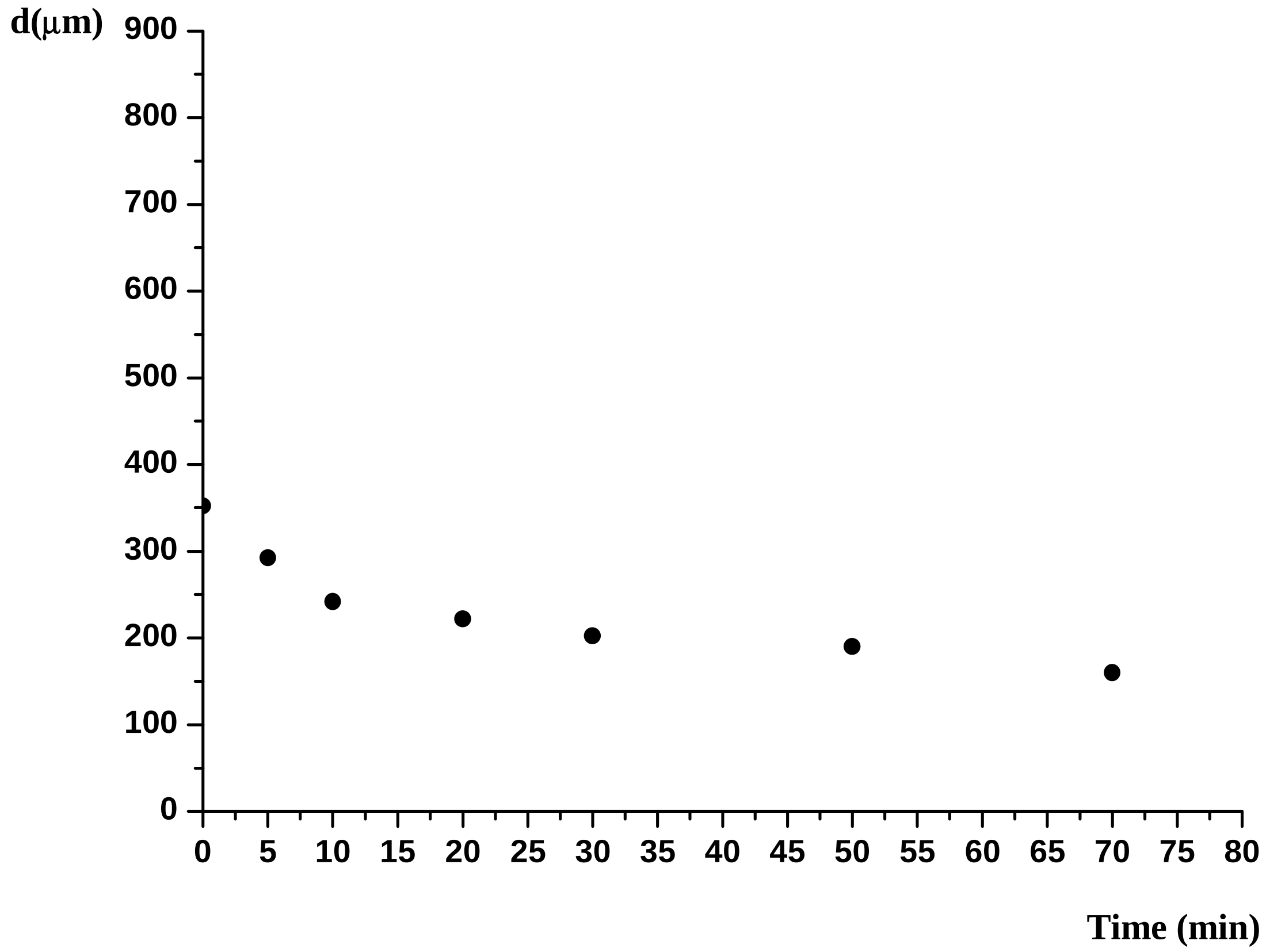
d(μm)
900
800
700
600
500
400
300
200
100
0
0 5 10 15 20 25 30 35 40 45 50 55 60 65 70 75 80
Time (min)

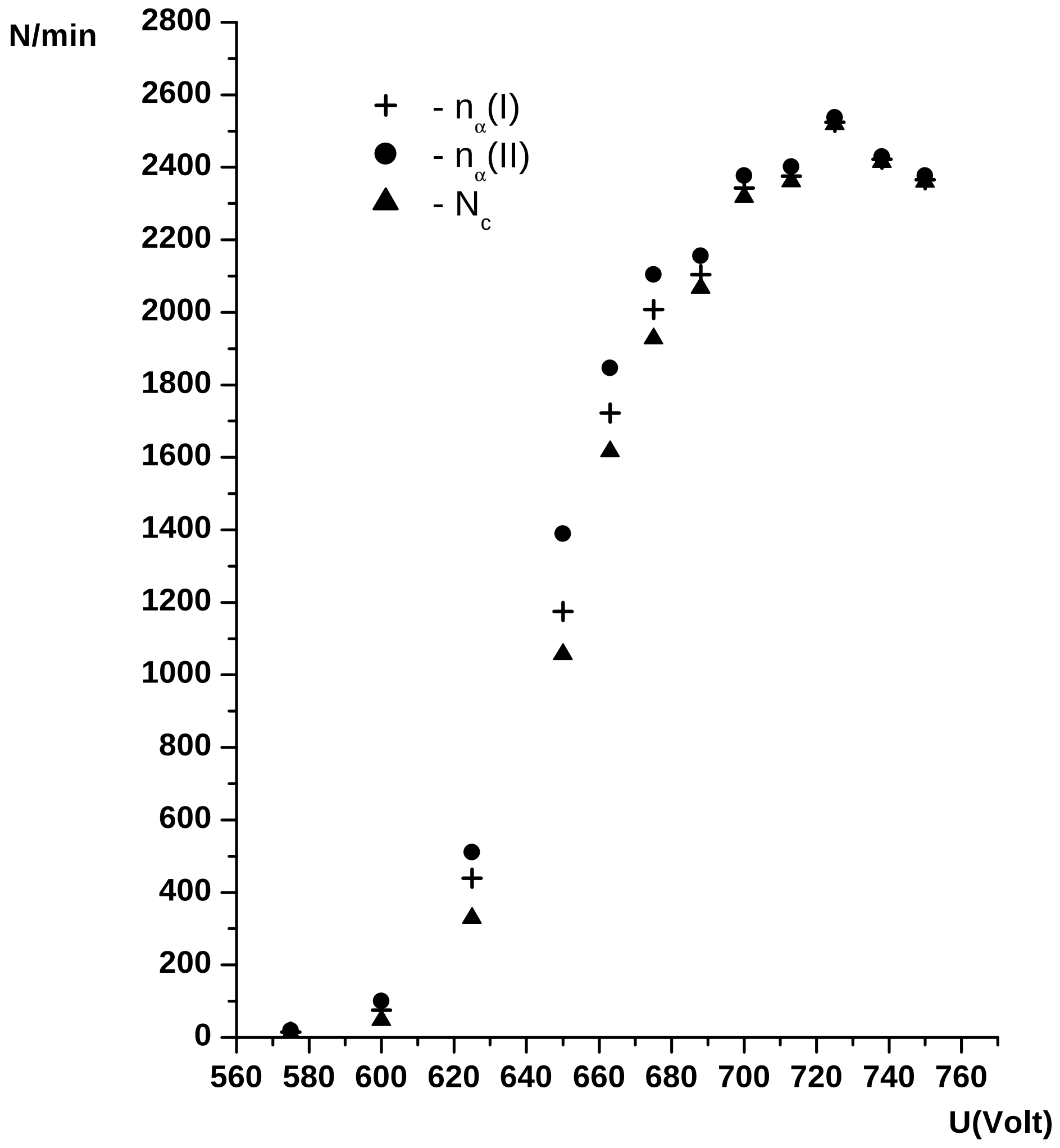
N/min
2800
2600
2400
2200
2000
1800
1600
1400
1200
1000
800
600
400
200
0
- $n_\alpha$(I)
- $n_\alpha$(II)
- $N_c$
560 580 600 620 640 660 680 700 720 740 760
U(Volt)

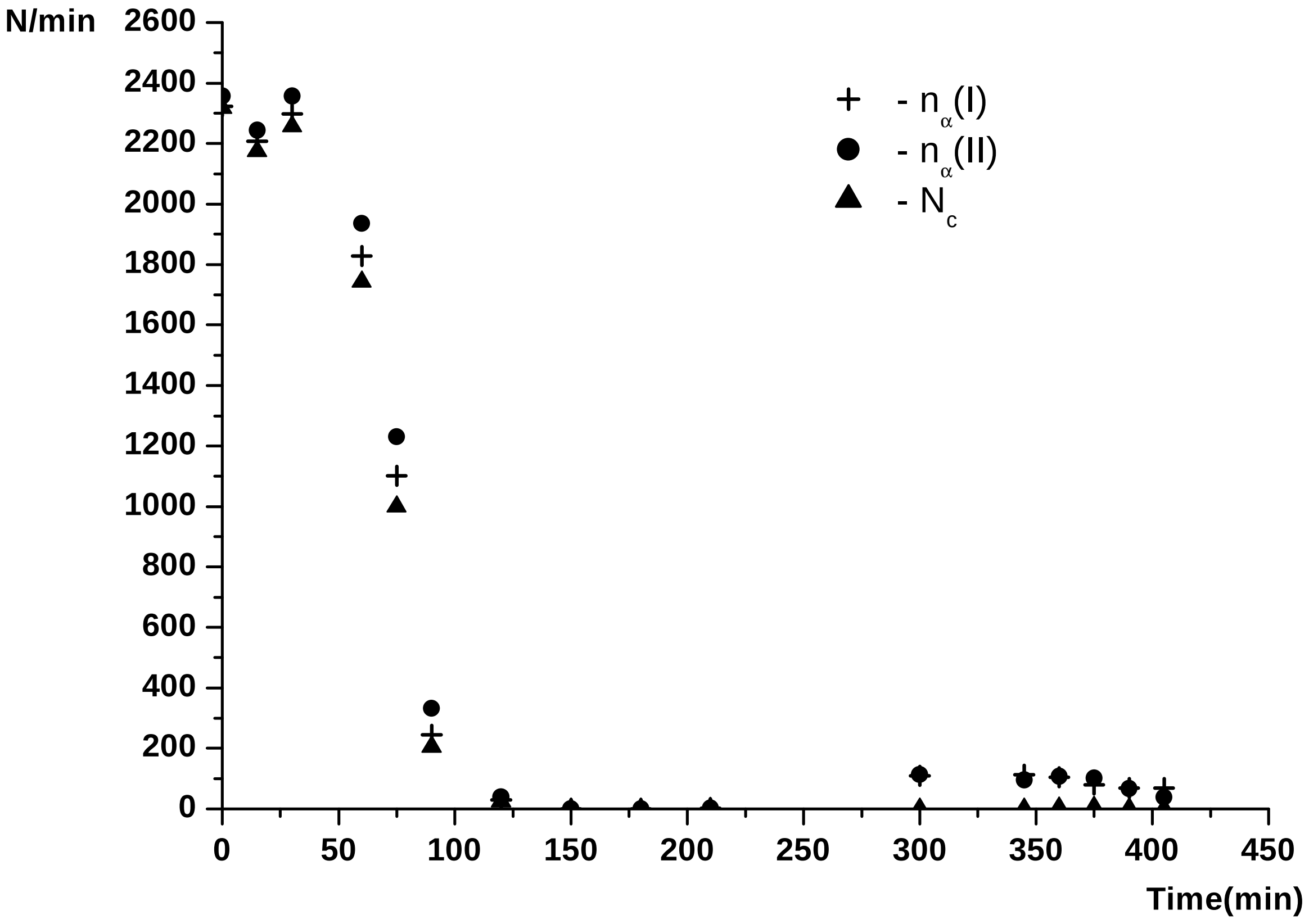

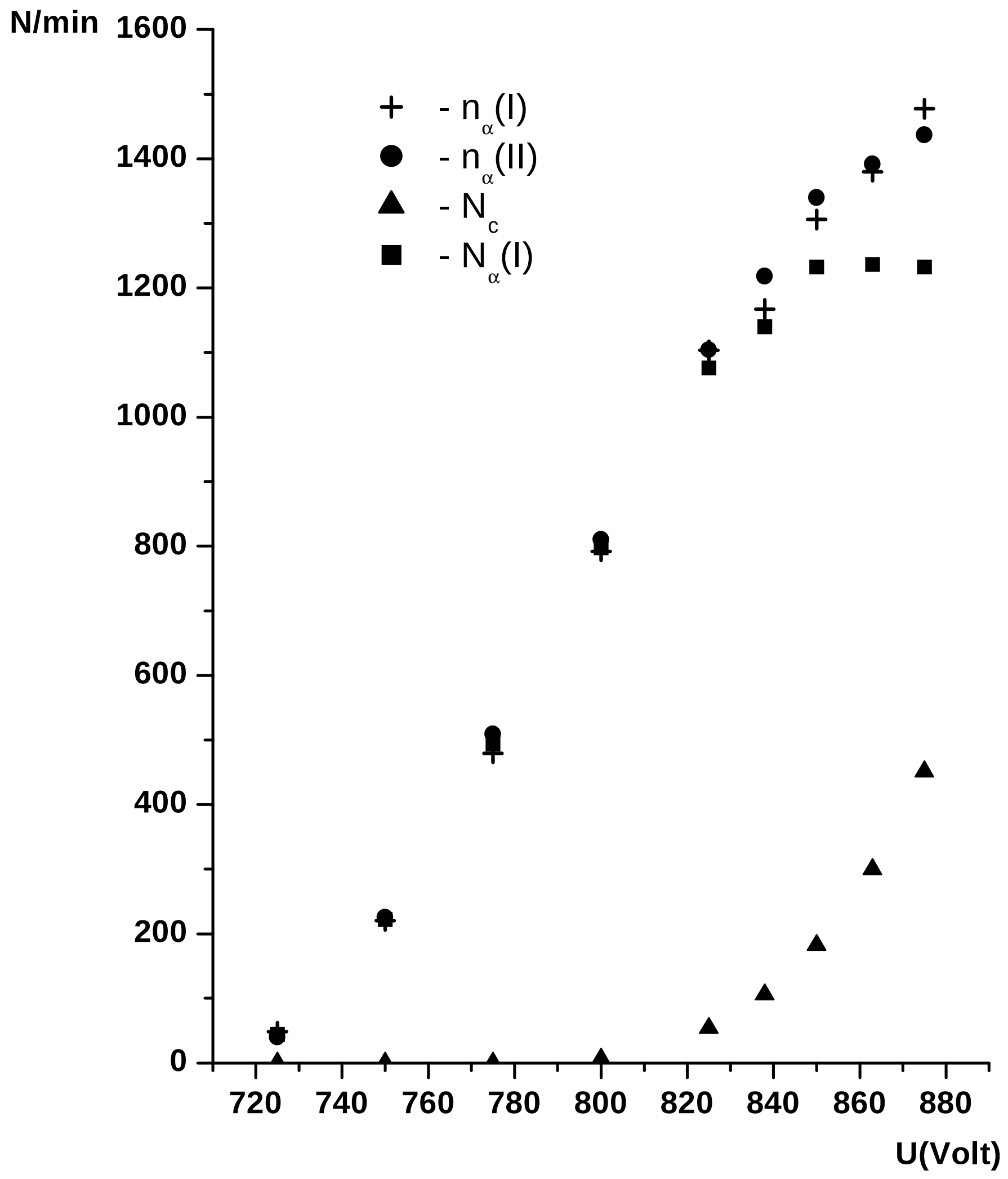

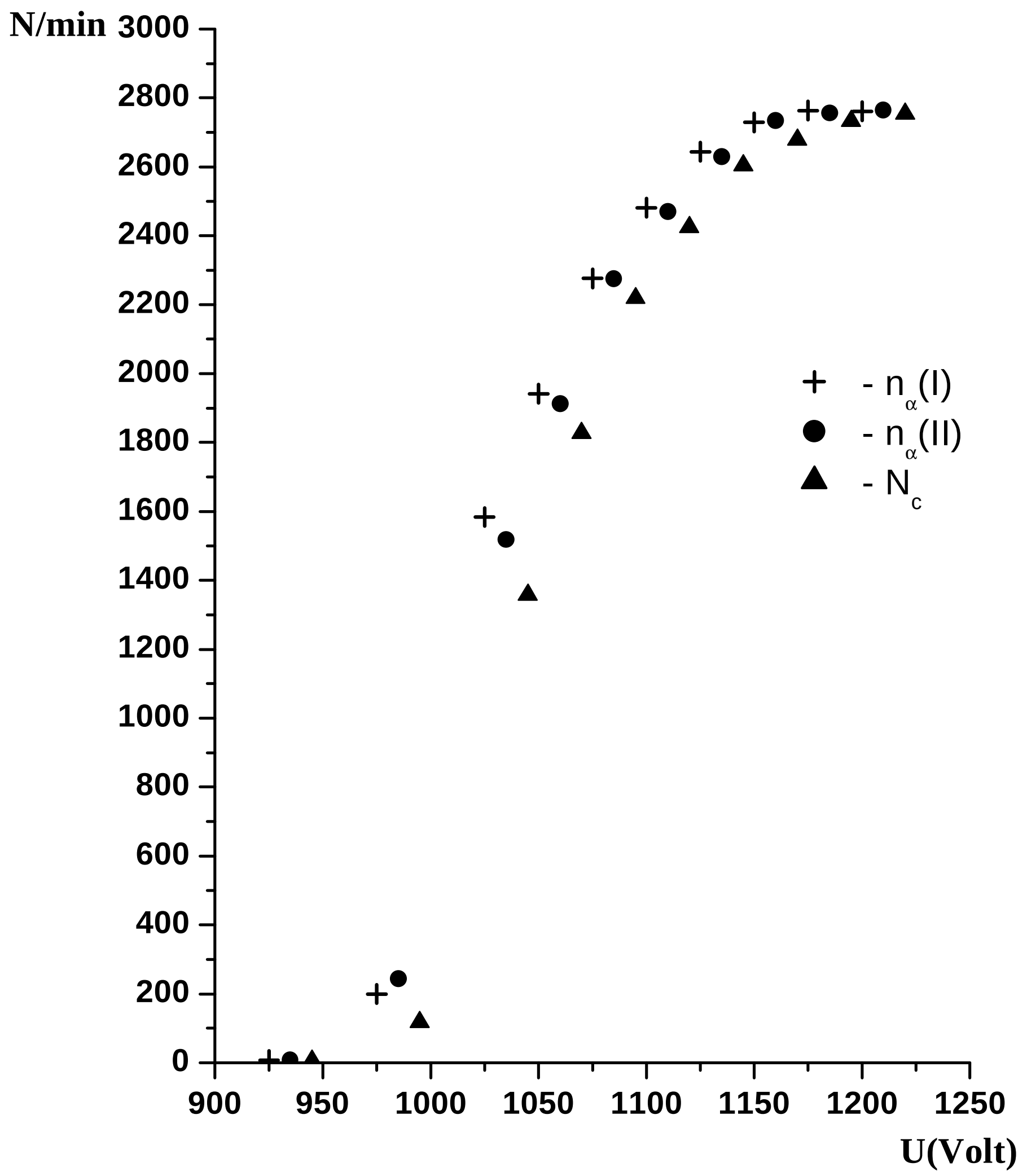

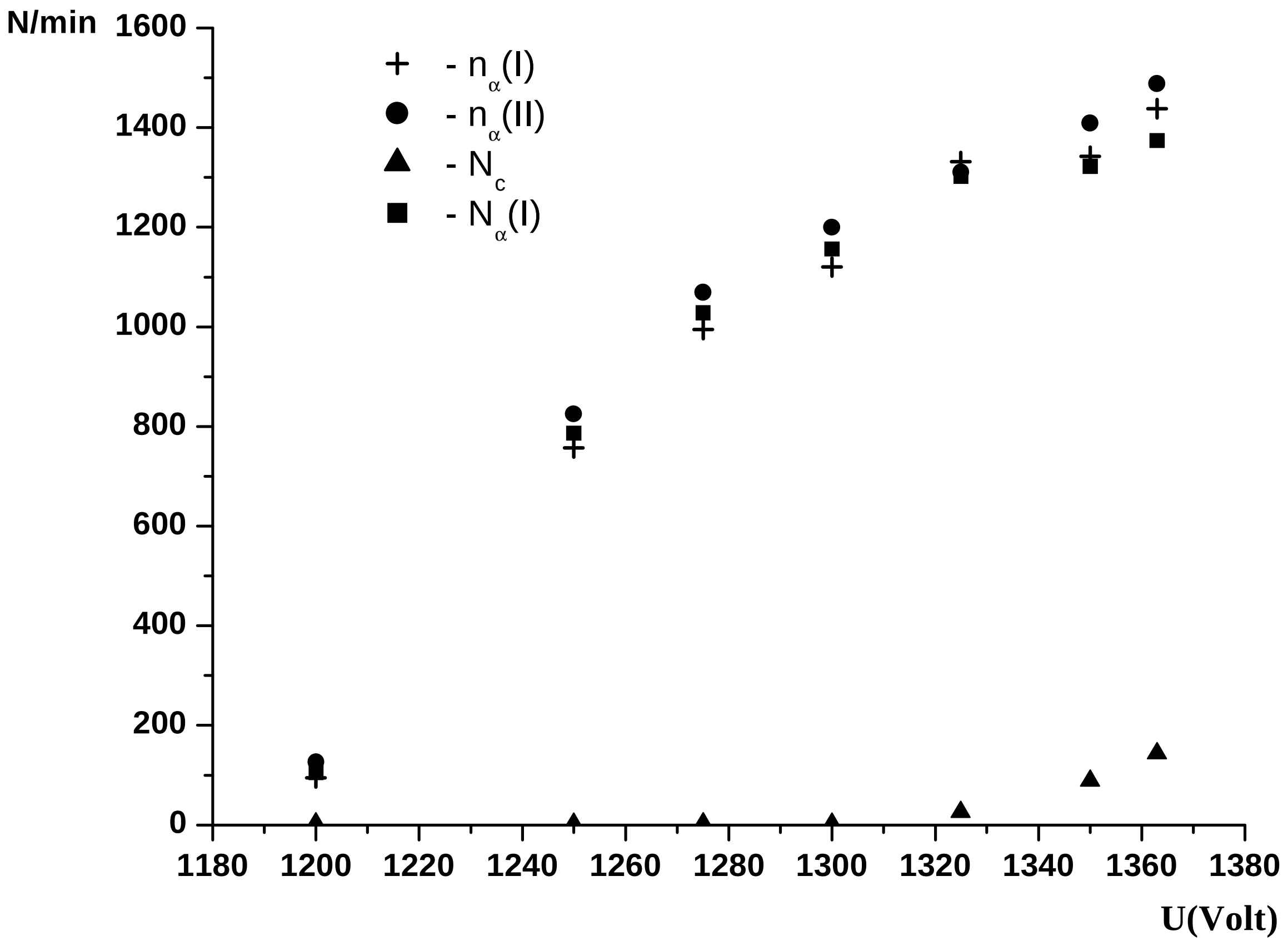

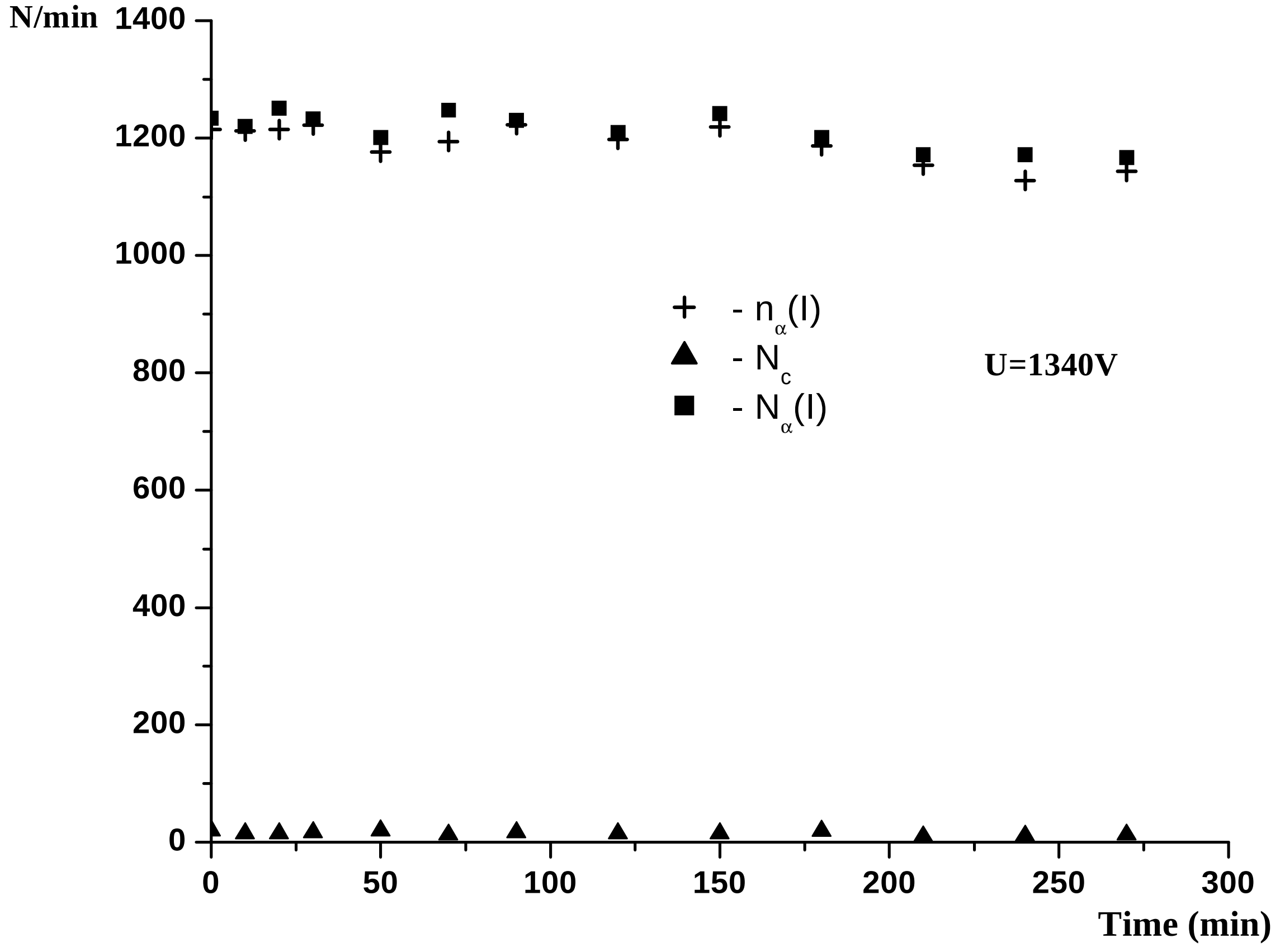

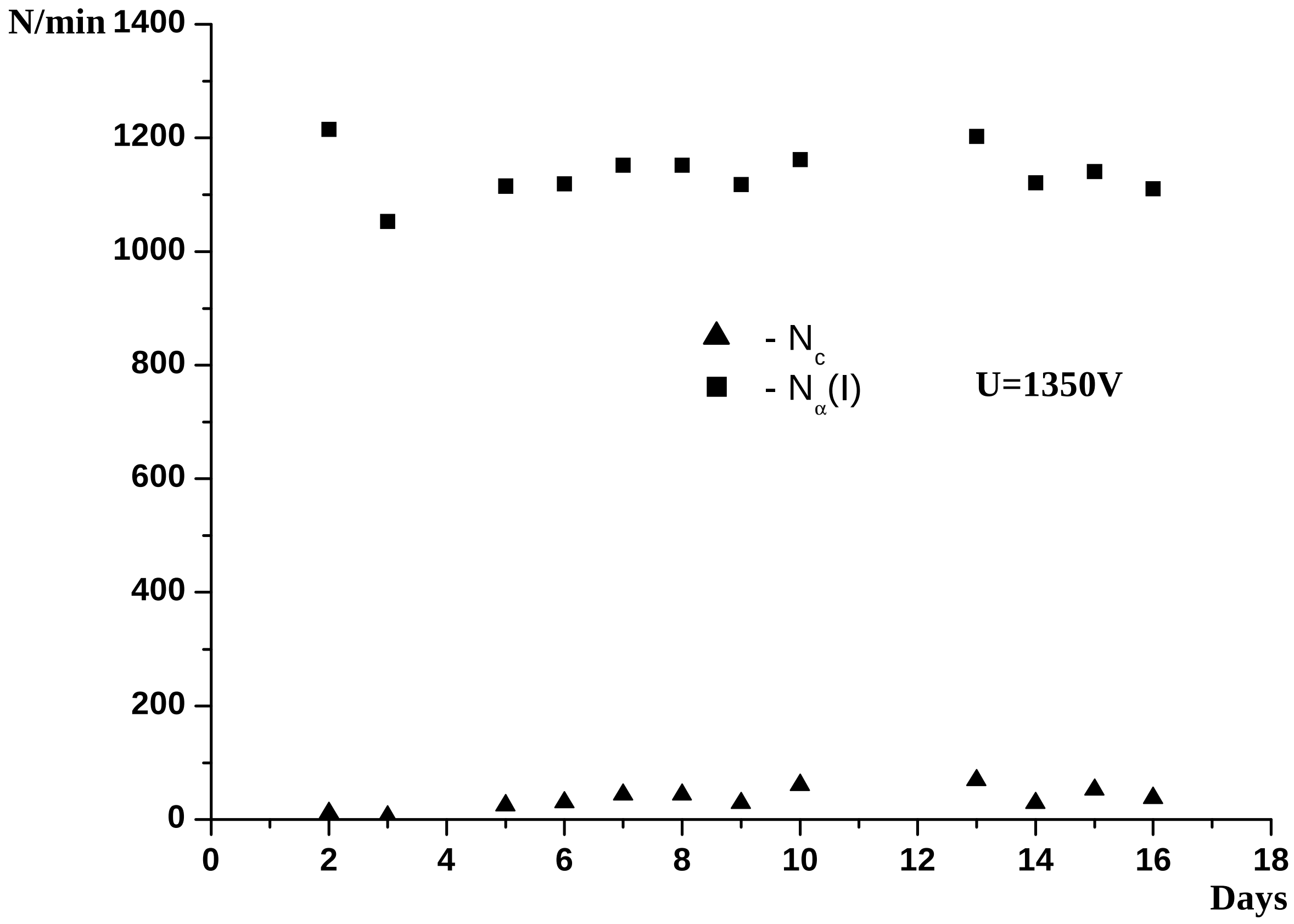

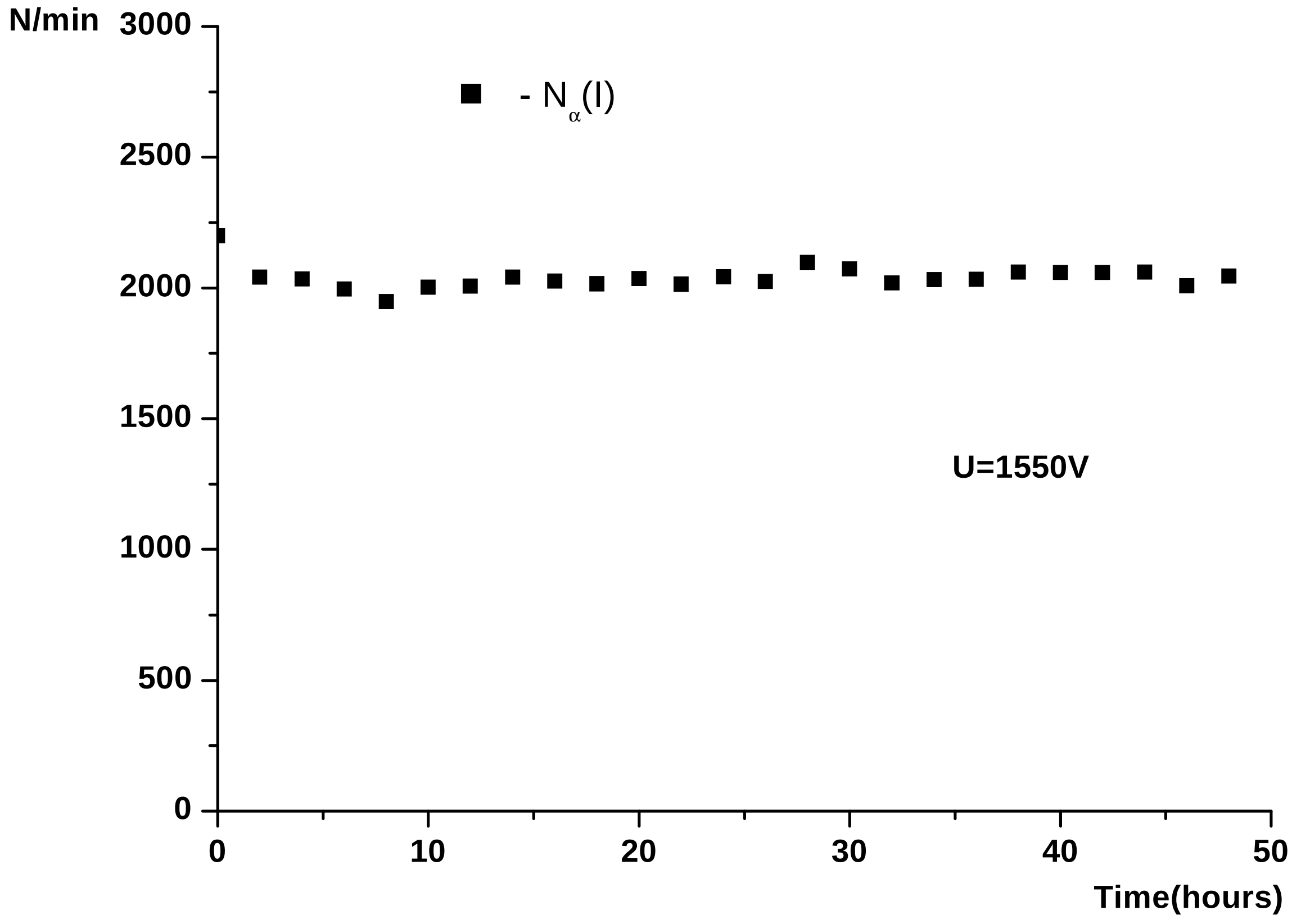

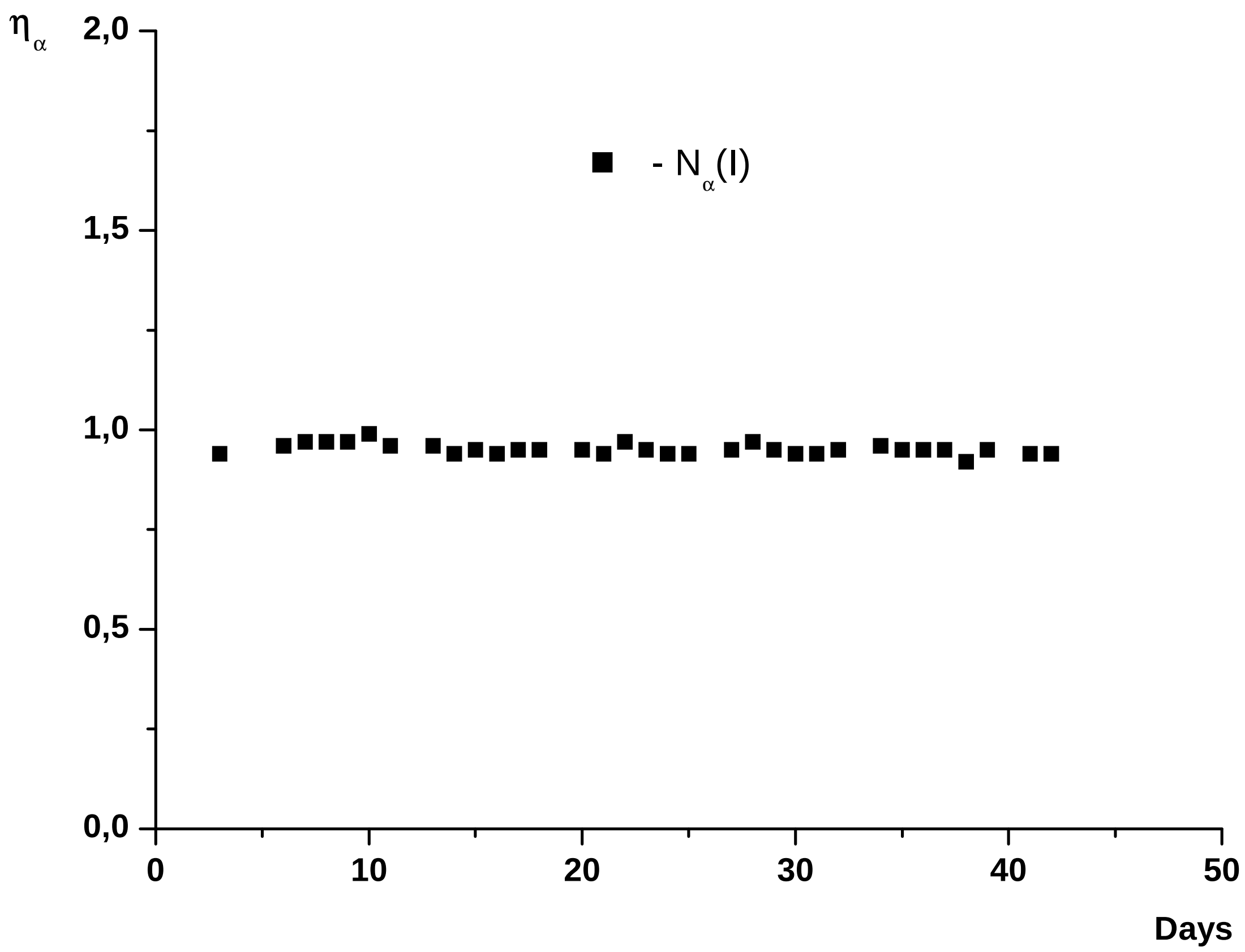

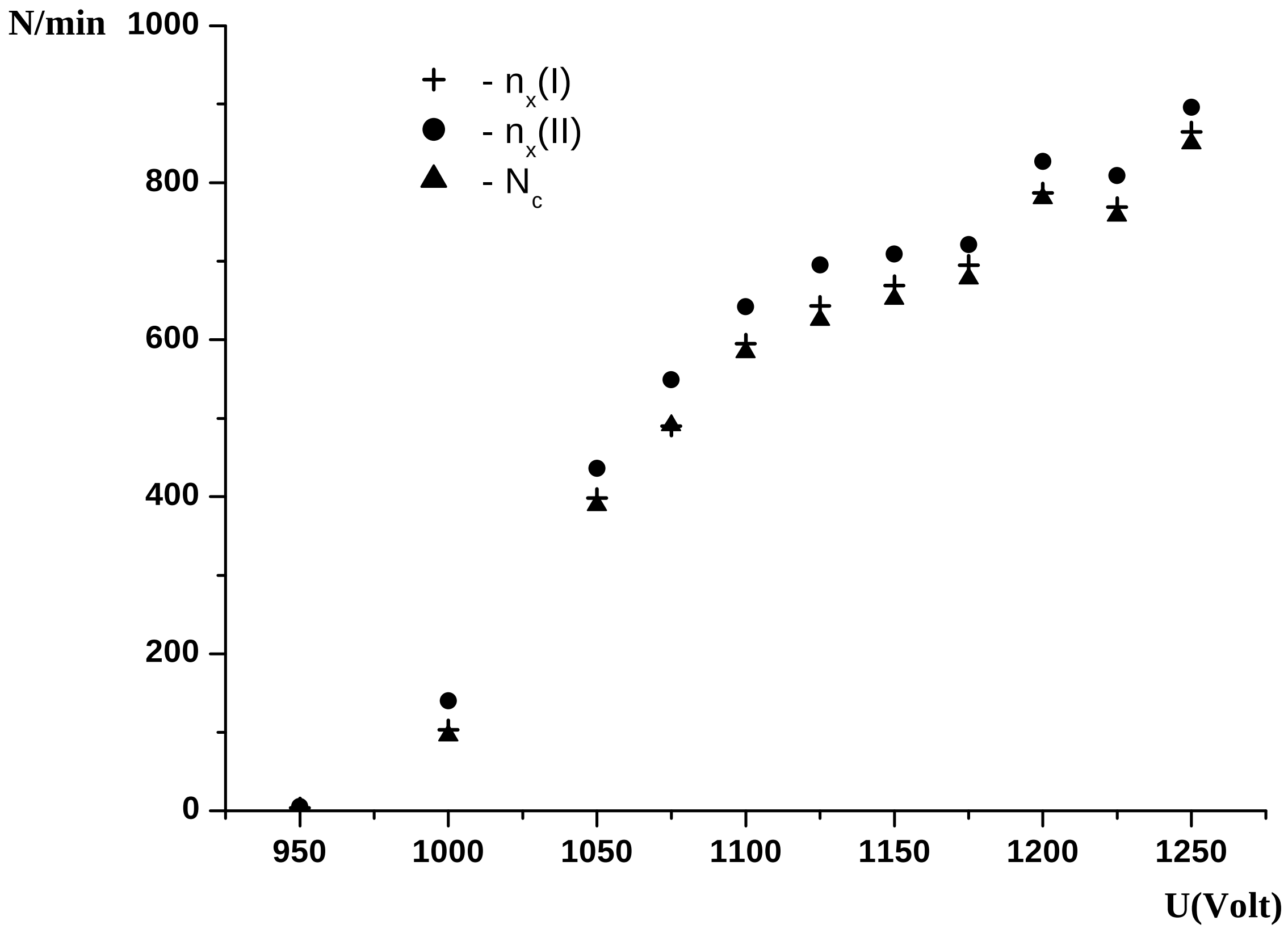

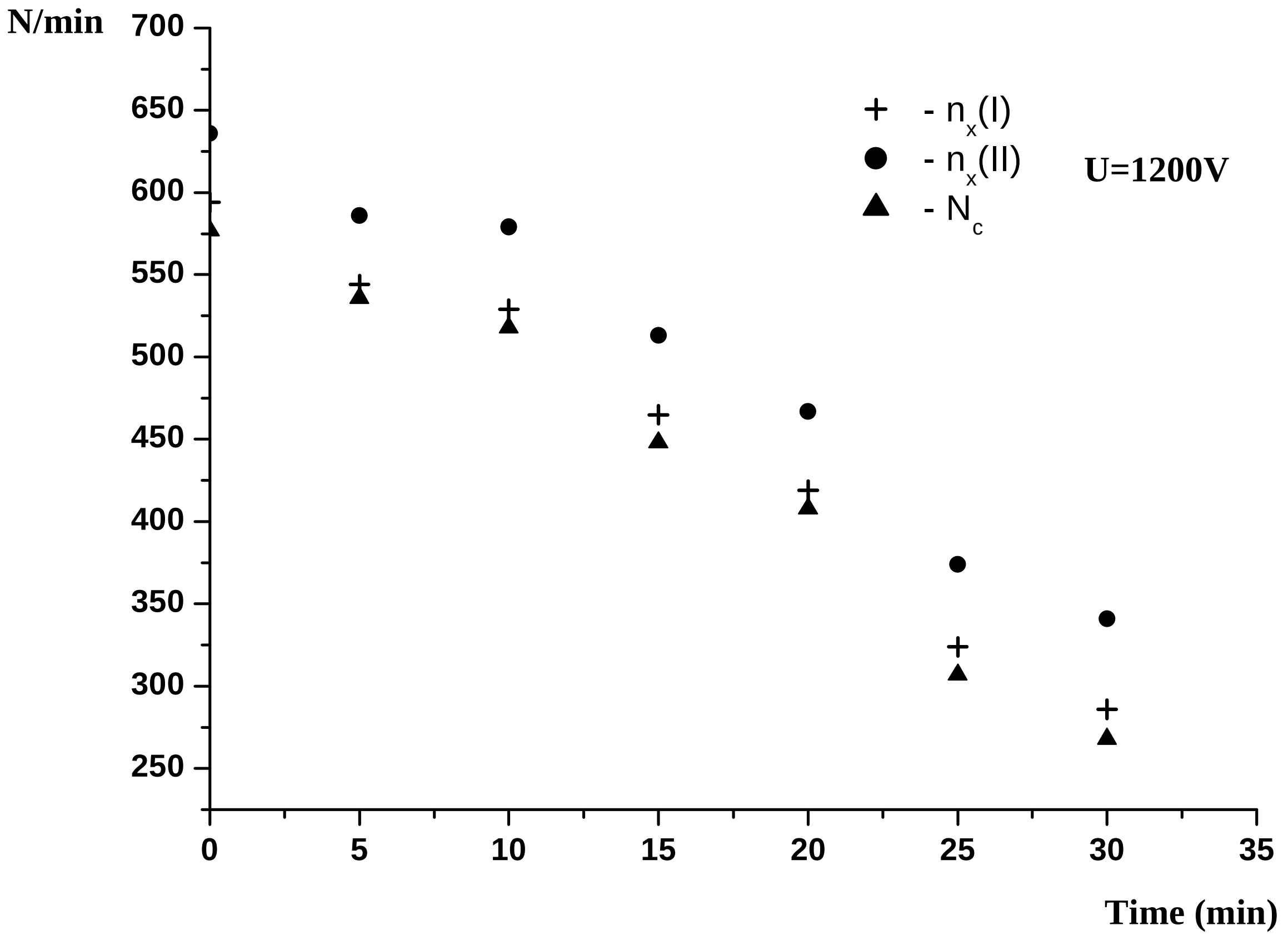

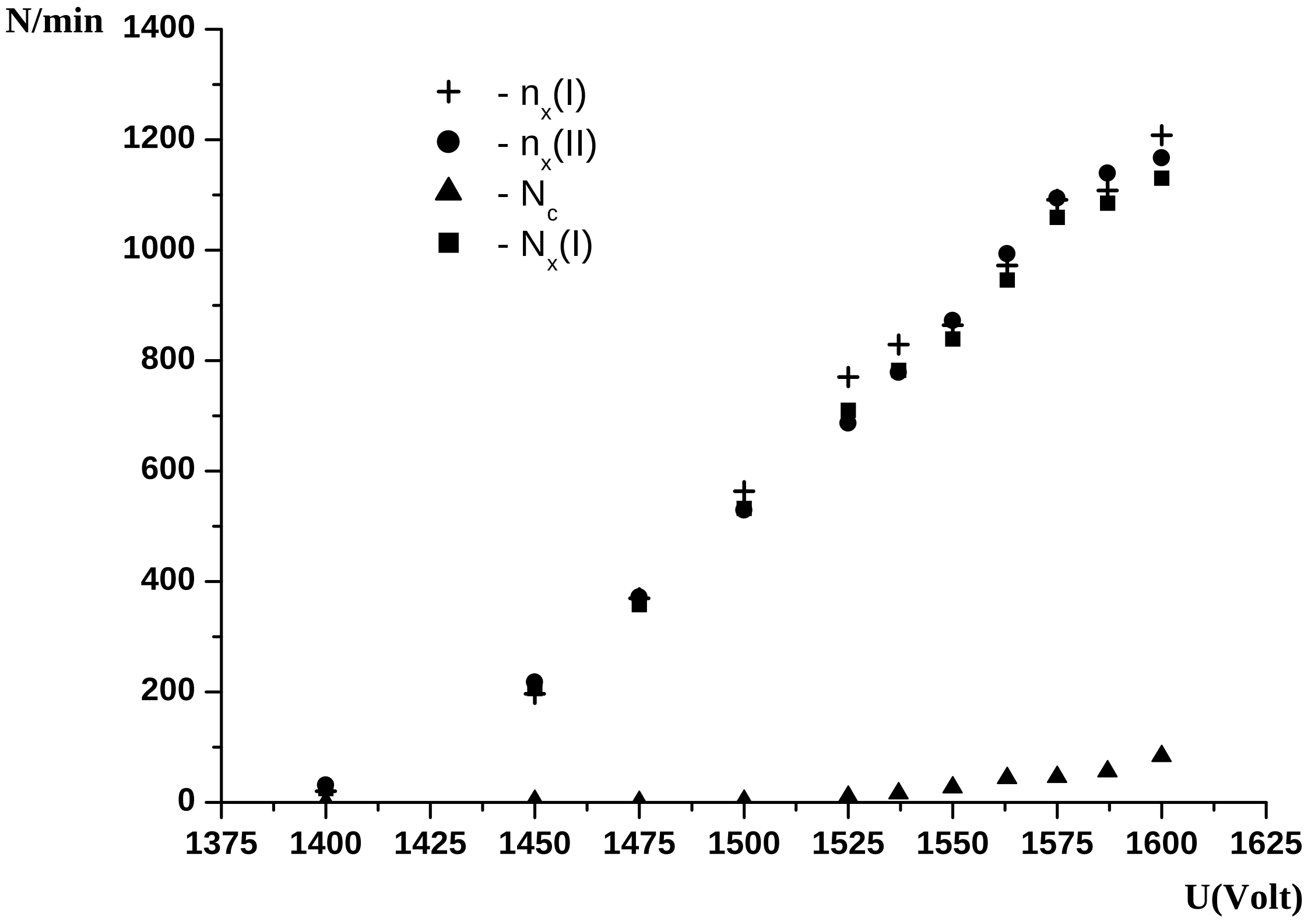

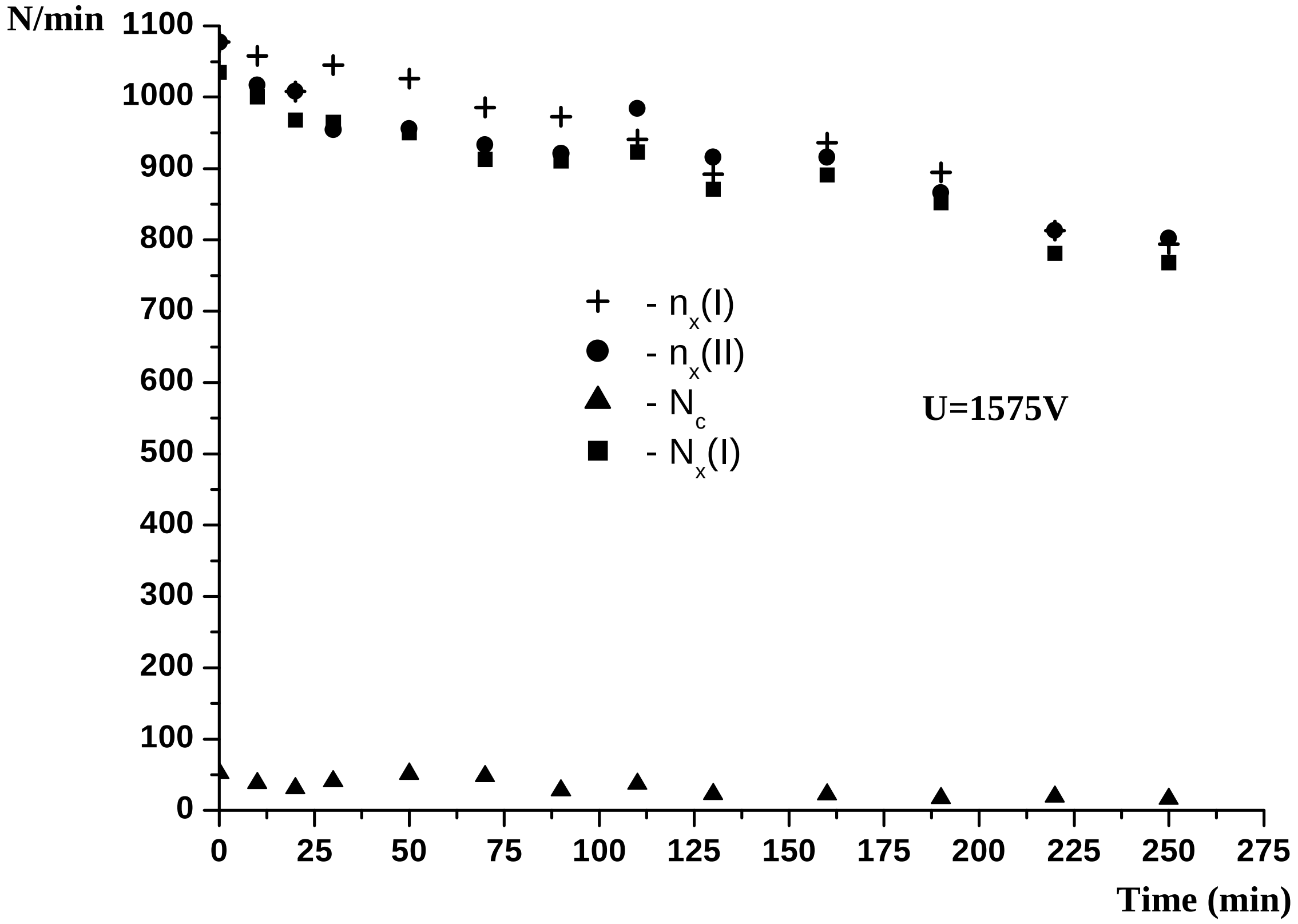

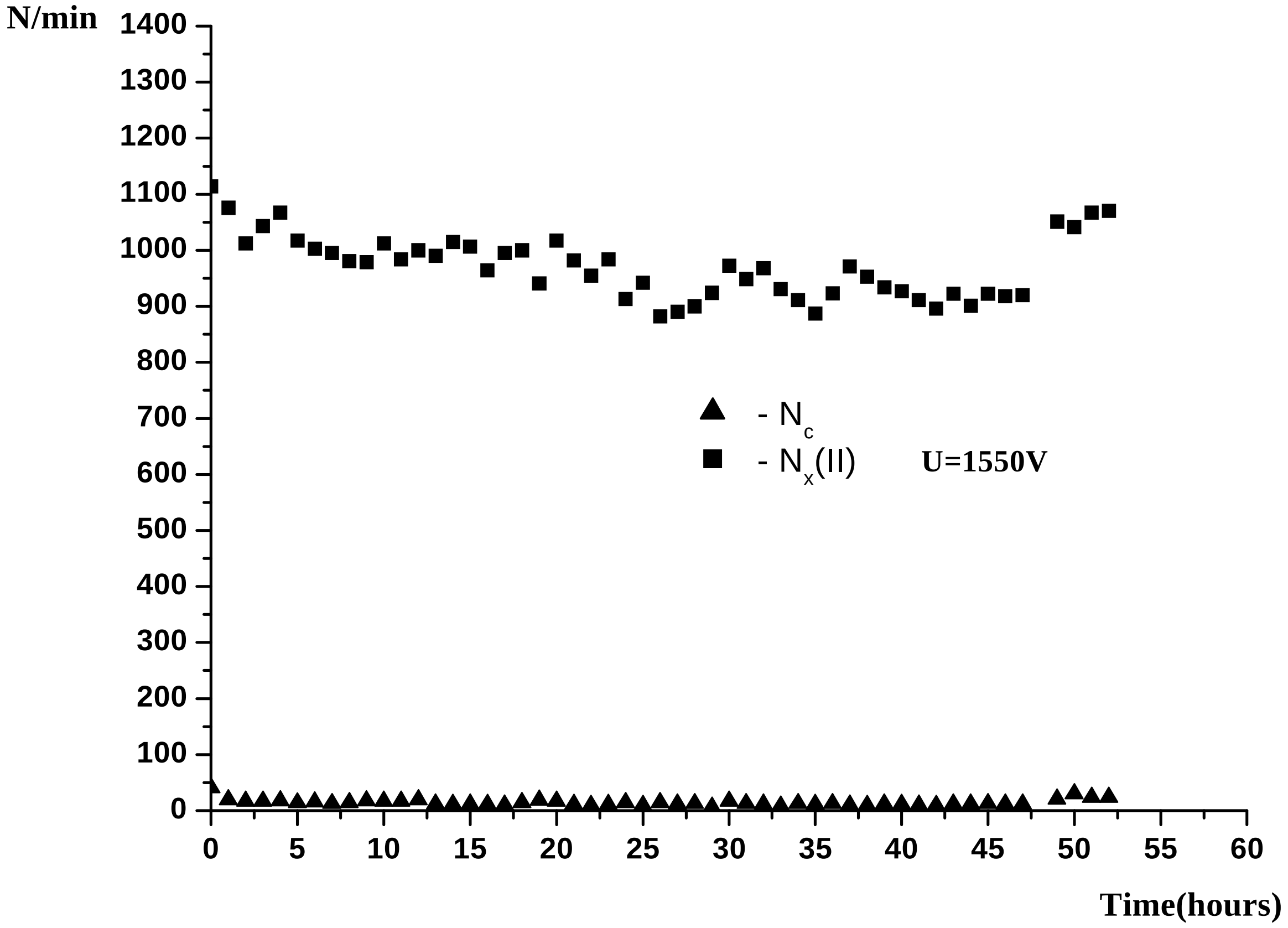

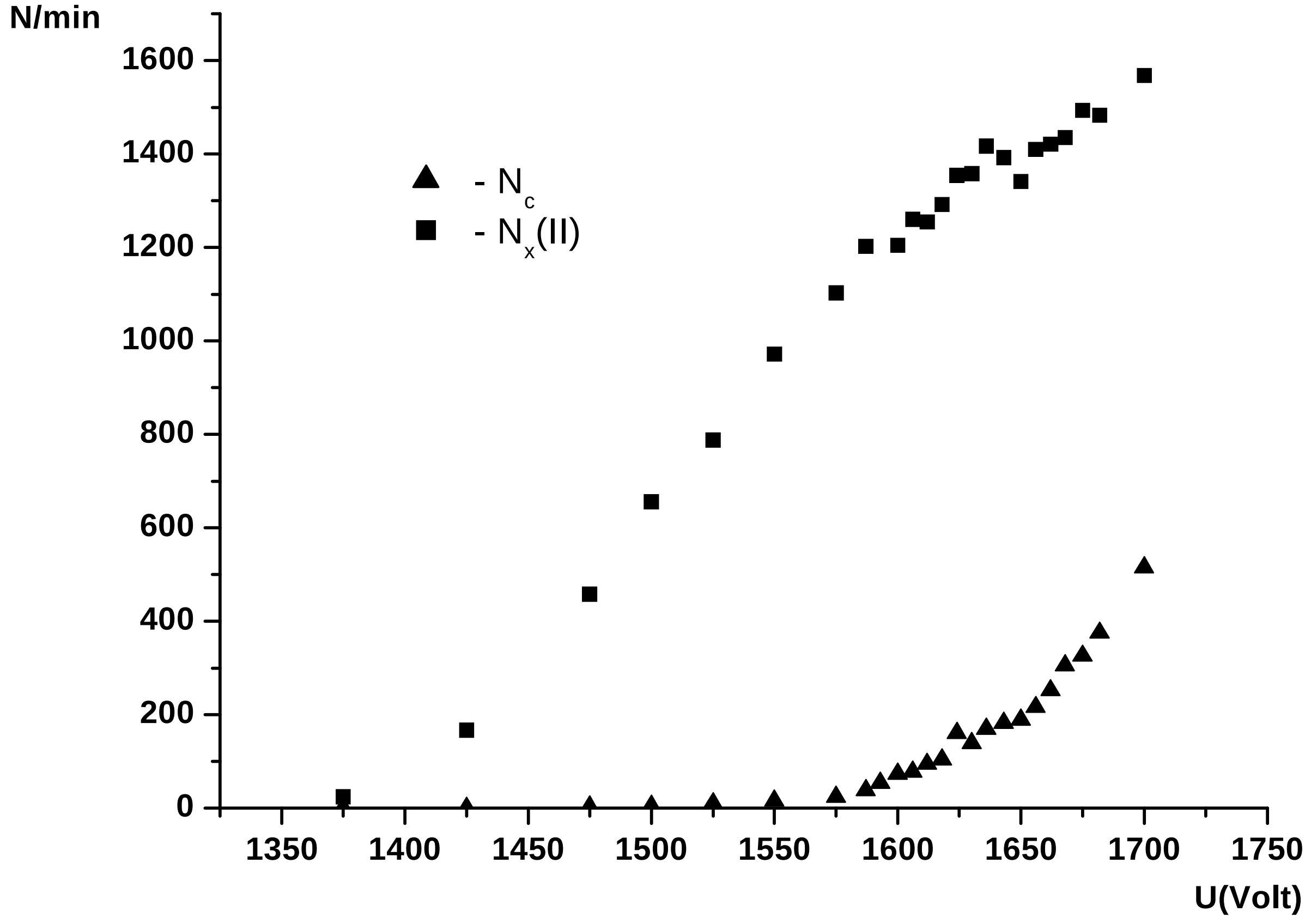